\documentclass[prl,letterpaper,twocolumn,showpacs,superscriptaddress]{revtex4}
\usepackage{graphicx,psfrag,amsmath,amssymb,amsfonts,bbm,latexsym,color,dcolumn}

\begin{document}

\title{Quantum two level systems and Kondo-like traps as possible sources of
decoherence in superconducting qubits.}
\author{Lara Faoro}
\author{Lev B. Ioffe}
\affiliation{Department of Physics and Astronomy, Rutgers University, 136 Frelinghuysen
Rd, Piscataway 08854, New Jersey, USA}
\date{\today }

\begin{abstract}
We discuss the origin of decoherence in Josephson junction qubits. We find
that two level systems in the surrounding insulator cannot be the dominant
source of noise in small qubits. We argue that electron traps in the
Josephson barrier with large Coulomb repulsion would give noise that agrees
both in magnitude and in temperature dependence with experimental data.
\end{abstract}

\pacs{85.25.Cp, 03.65.Yz,73.23.-b}
\maketitle



Despite remarkable experimental breakthroughs of recent years \cite%
{Nakamura1999,Nakamura2002,Vion2002,Martinis2002,Chiorescu2003,Wallraff2004,Pashkin2003,Chiorescu2004}%
, the realization of quantum computer based on Josephson elements remains far
away because developed circuits do not satisfy the stringent requirements on
the amount of allowed decoherence imposed by the quantum computation. Thus,
it is crucial to identify (and possibly eliminate) the microscopic origin of
the noise in these devices.

It is commonly believed that the most important source of noise, at least
for smaller qubits, is the fluctuations of the electric field generated by
the environment that are usually described in terms of the effective charge
induced on a superconducting island. Recent experiments performed on single
electron transistor (SET) and superconducting qubits provide an almost
complete characterization of the charge noise power spectrum as follows.

At low frequency ${(f<10^{3}\;Hz)}$ the charge noise spectrum of ${%
Al-AlO_{x}-Al} $ SET shows a $1/f$ dependence: ${S_{q}(\omega )=\frac{\alpha
^{2}}{\omega }}$ with intensity ${\alpha =10^{-3}-10^{-4}e}$ \cite%
{Zimmerli1992,Visscher1995,Zorin1996}. This level of the low frequency noise
agrees well with the observed dephasing (free induction decay) in
superconducting qubits \cite{Vion2002,Nakamura1999,Martinis2002}. Because
its effect is greatly reduced by echo techniques \cite{Nakamura2002}, it is
commonly believed that $1/f$ noise does not extend above $f_{M}\sim 1\;MHz$.
Recently, the direct measurements \cite{Oleg,Wellstood2004} of the
temperature dependence of the $1/f$-coefficient have shown the $T^{2}-$
dependence observed previously in different setups \cite%
{Wellstood1998,Kenyon2000}. At high frequencies ($f\gtrsim 1\;GHz$) the
spectrum and the decay rate of the excited state of the superconducting
qubits indicate the presence of a few quantum two level systems (TLSs)
strongly coupled to the qubit \cite{Martinis2002,Astafiev2004} superimposed
on the Ohmic power spectrum \cite{Astafiev2004}. Noteworthy, the crossover
frequency between the high frequency linear noise and the interpolated low
frequency $1/f$ noise is of order of $T $, which suggests the same physical
origin of the low and high frequency noise \cite{Astafiev2004}.

In this Letter we discuss two microscopic mechanisms that might be
responsible for the charge noise. First, we consider the effect of weakly
interacting TLSs hosted in the amorphous material surrounding
superconducting islands. We prove that this environment provides a
significant source of decoherence but the detailed characteristics of its
noise power spectrum are in qualitative and quantitative disagreement with
the data; in particular, the observed $T^2$ dependence of $\alpha^2$
requires unphysical density of states of TLSs \cite{Shnirman2005}. Second,
we discuss the effect of Kondo-like traps located inside the insulating
Josephson contact. We show that these traps behave similarly to weak Andreev
fluctuators. \cite{Faoro2004}. We argue that for realistic parameters these
traps produce noise very similar to the one observed experimentally.

\textit{Quantum TLSs model.} It is well established that TLSs dominate the
low temperature behavior of most amorphous materials at low temperatures 
\cite{Black-Halperin,Burin1995,Levitov1990}. One usually assumes that these
TLSs correspond to atoms that tunnel between two positions. Although their
exact microscopic origin is not known, it is well established that they are
well described by a pseudospin characterized by dipole moment $\widehat{p}=%
\overrightarrow{p}\sigma ^{z}=e\vec{d}\sigma ^{z}$ and controlled by the
Hamiltonian 
$H_{0}=E\sigma _{z}+t\sigma _{x}$. \ 
Here $E$ is the energy difference between the two minima, $t$ is the
tunneling between them and those are distributed according to: 
\begin{equation}
P(E,t)=\frac{\nu }{t}~~~~~~\frac{\nu e^{2}d^{2}}{\epsilon }\sim 10^{-3}
\label{glass}
\end{equation}%
where ${\nu =10^{20}/cm^{3}eV}$ is the typical density of states of TLSs, $d$
is the atomic distance and $\epsilon $ denotes the dielectric constant of
the medium. Different TLSs interact with each other via dipole-dipole
interactions: 
\begin{equation}
H_{int}=\sum_{i}\sum_{j\neq i}\frac{\hat{p}_{i}\hat{p}_{j}-3(\hat{r}_{ij}%
\hat{p}_{i})(\hat{r}_{ij}\hat{p}_{j})}{4\pi \epsilon r_{ij}^{3}}\;
\label{dipint}
\end{equation}%
where $r_{ij}$ is the distance between the $i^{th}$ and $j^{th}$ dipole.
Notice that the effective strength of the interactions is given by the
dimensionless parameter (\ref{glass}) and is always very weak. This, for
instance, implies that the effective field that other thermally active
dipoles produce on a given one is small compared to the temperature. This
small fluctuating field has, however, an important effect on the coherent
dynamics and is responsible for the low frequency $1/f$ noise, as we show
below.

In order to estimate the coupling between a TLS and the qubit we notice
that, in the charge basis of the qubit, its two states are characterized by
the electrostatic potential operator $\delta \mathcal{V}\sigma _{qub}^{z}$
and thus produce electric fields that differ by ${\delta \mathcal{E}=\delta 
\mathcal{V}/L}$ at the position of the dipole, leading to the qubit-dipole
interaction ${H_{d-qub}=\delta \mathcal{E}\widehat{p}\sigma _{qub}^{z}}$.
The characteristic length $L$ is determined by the dipole position: it is of
the order of the barrier thickness, ${3\;nm}$, for the dipoles in the
barrier or ${L\approx 500\;nm}$ for the dipoles in the surrounding
insulator. The resulting effect on the qubit is conventionally described as
the induced charge $\delta Q=\kappa e\,\sigma ^{z}$ where the coupling
constant ${\kappa =p/eL\sim d/L}$. Note that the coupling constant with the
dipoles in the barrier is rather large, $\kappa \sim 10^{-1}$.

Because the interaction term between the dipoles is small, it is convenient
to work in the basis that diagonalizes the Hamiltonian $H_{0}$. In this
basis ${H_{0}=\sum {\Delta E_{j}}\tau _{j}^{z}}$ where ${\Delta E_{j}=\sqrt{%
E_{j}^{2}+t_{j}^{2}}}$ . The interaction between the dipoles has two
effects: it causes level broadening ($1/T_{2}$) and it gives each level a
finite decay time even in the absence of phonons; we consider these effects
in turn. We begin with level broadening. For this calculation we can neglect
the back reaction and assume that dipoles fluctuate independently. In this
approximation each dipole is subjected to the effective field $h(t)$
generated by surrounding independent TLSs: 
\begin{eqnarray}
H_{int}^{eff} &=&\sum_{i}h_{i}(t)\bigl (\cos \theta _{i}\tau _{i}^{z}-\sin
\theta _{i}\tau _{i}^{x}\bigr )  \label{dipinteff} \\
\;\langle \langle h_{i}^{2}\rangle \rangle &=&\sum_{j\neq i}k_{ij}^{2}\cos^2
\theta _{j}\frac{1}{\cosh ^{2}\beta \Delta E_{j}}  \label{brod}
\end{eqnarray}%
where the coupling ${k_{ij}=\frac{\vec{p}_{i}\vec{p}_{j}-3(\hat{r}_{ij}\vec{p%
}_{i})(\hat{r}_{ij}\vec{p}_{j})}{4\pi \epsilon r_{ij}^{3}}}$. Averaging (\ref%
{brod}) over the distribution (\ref{glass}) we obtain the typical level
broadening: ${\ \gamma \approx \left( \frac{\nu p^{2}}{\epsilon }\right) T}$%
. As expected, the dephasing level broadening is always much less than the
temperature. Note that this estimate assumes a significant total number of
thermally activated dipoles; this condition might not be satisfied in the
thin Josephson barrier.

The relaxation rate $\gamma ^{1}$ of dipoles is due to phonon emission or to
the excitation of another dipole. The rate of the former decreases very
rapidly at low energies ${\gamma ^{1}\sim \Delta E^{3}/\omega _{D}^{2}}$
where $\lambda_D \sim 10^{2}$ and becomes very low ($\sim 1\;ms$) for $%
\Delta E<50\;mK$. The latter process dominates at very low temperatures. The
time average of the relaxation rate of a given dipole can be found from
Fermi Golden Rule: 
\begin{equation}
\gamma _{i}^{1}=\sum_{j}{}^{\prime }\frac{\gamma }{\left( \Delta
E_{i}-\Delta E_{j}\right) ^{2}+\gamma ^{2}}|k_{ij}|^{2}\sin ^{2}\theta
_{i}\sin ^{2}\theta _{j}\;  \label{relax}
\end{equation}%
This neglect of the higher order terms in the interaction can be justified
if the dipole-dipole interaction is weak compared to the line width $%
(\left\vert k_{ij}\right\vert <\gamma )$: more strongly coupled dipole pairs
form new effective dipole modes and thus have to be excluded from the sum
which is indicated by prime. Averaging (\ref{relax}) we get for thermally
excited dipoles:\, ${\gamma _{i}^{1}\approx \sin ^{2}\theta _{i}\left( \frac{%
\nu p^{2}}{\epsilon }\right) ^{2}T}$ \cite{Lev1}.

As explained above, the effect of the fluctuating dipoles on the qubit is
characterized by the induced charge noise spectrum: 
\begin{equation}
S_{q}(\omega )=\int_{-\infty }^{\infty }dt \langle \delta Q(t) \delta
Q(0)\rangle e^{i \omega t} = \left (\kappa e \right )^2 G(\omega).\;
\label{spectrum}
\end{equation}
with correlator $G(\omega )=\sum_{i}\int_{-\infty }^{\infty }dt \langle
\sigma_{z}^{i}(t)\sigma _{z}^{i}(0)\rangle e^{i\omega t}.$

At high frequency the correlator is dominated by the resonant dipoles giving 
\begin{equation}
G(\omega )=\sum_{i}\sin ^{2}\theta _{i}\frac{\gamma }{(\omega -2\Delta
E_{i})^{2}+\gamma ^{2}}\;  \label{highf}
\end{equation}%
By substituting (\ref{highf}) into (\ref{spectrum}) and by averaging over
the probability distributions we get 
\begin{equation}
S_{q}(\omega )=\left[ \left( \frac{\nu p^{2}}{\epsilon }\right) \frac{%
\epsilon V}{e^{2}L^{2}}\right] e^{2}\;  \label{npsh}
\end{equation}%
where $V$ is the volume of the amorphous insulator.

We conclude that the fluctuating dipoles give white noise spectrum at high
frequencies. To estimate the noise level we first check that the amount of
amorphous substance is sufficient to provide a large number of thermally
activated TLSs. In real qubits, TLSs are located in the barrier or in the
surrounding oxide. For the barrier we estimate ${V\approx 10^{7}\mathring{A}%
^{3}}$, so that the dipole density of states is about $0.1/K$. Clearly this
is not enough to give a continuous spectrum and we must conclude that the
barrier can contribute to the charge noise at high frequency by providing
only few TLSs, strongly coupled to the qubit; this might at most account for
resonance peaks. The situation is different for the surrounding oxide which
has volume $V\gtrsim 10^{9}A^{3}$, in this case the number of thermally
activated TLSs might be sufficient to produce a continuous spectrum of
relaxation times with $S_{q}(\omega )\approx 10^{-6}e^{2}\;K^{-1}$.
Experimentally, the linear high frequency noise and the low frequency $1/f$
noise both extrapolate to the same value $\ \sim {10^{-6}e^{2}}\;K^{-1}$ at $%
\omega \sim T\approx 120\;mK$ \cite{Astafiev2004}. This value is very close
to our estimate showing that the TLSs might provide a non-negligible source
of noise at these intermediate frequencies.

At low frequencies the noise is dominated by thermally excited dipoles that
behave as classical fluctuators at frequencies less than their relaxation
rates{: 
\begin{equation}
G(\omega )=\sum_{i}\cos ^{2}\theta _{i}\frac{2\gamma _{1}^{i}}{\omega
^{2}+(\gamma _{1}^{i})^{2}}\;  \label{lowf}
\end{equation}%
The relaxation rates }$\gamma _{1}^{i}$ have a wide distribution, ${%
P(\gamma)\sim 1/\gamma} $, producing upon averaging, $1/f$ noise: 
\begin{equation}
S_{q}(\omega )=\left( \frac{\nu p^{2}}{\epsilon }\right) \left( \frac{
\epsilon V}{e^{2}L^{2}}\right) \frac{T}{\omega }e^{2}\;  \label{dip1fnoise}
\end{equation}%
Its linear temperature dependence is a direct consequence of the constant
density of states that results in a linear dependence of the number of
thermally activate TLSs: ${n_{TLS}=\frac{T}{W}\rho _{0}V}$ where $W\approx
1eV $ is the energy scale of the disorder and $\displaystyle{\rho
_{0}=10^{20}/cm^{3}}$ is the bare density. Similarly to the high frequency
noise, this noise is of the right order of magnitude at lowest temperatures
but grows linearly with temperature instead of following the $T^{2}$
dependence. Moreover, the phenomenological assumption that the number of
thermally active dipoles is ${n_{TLS}=\frac{T^{2}}{W^{2}}\tilde{\rho}_{0}V}$
has an other problem: the observed value of $1/f$ noise at low frequency
implies that the total density ${\tilde{\rho}_{0}=\frac{W}{T}\rho
_{0}\approx 10^{6}\rho _{0}}$ is much larger than one per atom. Thus, in
order to account for the data, one needs a physical mechanism characterized
by a small energy scale $W$.

\textit{Kondo-like traps model.} A natural candidate for the small energy
scale is the superconductive gap $\Delta $. In the bulk of superconductor
all electron states are gapped and low energy electron traps seem unlikely.
It is very likely, however, that numerous electron traps appear in the
insulator at the SI boundary, with energies distributed in a broad energy
range. The Coulomb repulsion between two electrons in the same trap is very
strong prohibiting double occupancy. In the absence of such repulsion, the
effect of the electrons hopping from one trap to another and their hopping
via Andreev process would give the correct frequency and temperature
dependence \cite{Faoro2004}, but requires an unphysical density of states of
these traps, similar to the TLS discussed above. The strong on-site
repulsion leads to a formation of a weak Kondo-like resonance at the Fermi
surface and provides the desired low energy scale. This physics is described
by $H=H_{d}+H_{sd},$ where $H_{d}$ and $H_{sd}$ describe the Hamiltonian of
the Anderson-like traps and of the coupling between the quantum impurities
and the superconductor: 
\begin{eqnarray}
H_{d} &=&\sum_{i\sigma }\epsilon _{d}^{0}c_{di\sigma }^{\dagger }c_{di\sigma
}+U\sum_{i}n_{di\uparrow }n_{di\downarrow },\;  \notag \\
H_{sd} &=&\sum_{i}\sum_{k\sigma }\left( t_{ki}c_{k\sigma }^{\dagger
}c_{di\sigma }+h.c.\right) ,\;  \label{Kondo1}
\end{eqnarray}%
Here $n_{di\sigma }=c_{di\sigma }^{\dagger }c_{di\sigma }$, ${c_{k\sigma
}^{\dagger }}$ (${c_{di\sigma }^{\dagger }}$) is the creation operator of a
conduction (localized) electron and $U$ is the on-site repulsion energy. In
the absence of superconductivity, the Kondo effect \cite{Kondo} leads to the
appearance of the resonance at the Fermi energy with width ${%
T_{K}^{i}=\Gamma _{i}\sqrt{\frac{U}{2\Gamma _{i}}}\exp \left[ -\pi \frac{%
\epsilon _{d}^{0}}{2\Gamma _{i}}\left( 1+\frac{\epsilon _{d}^{0}}{U}\right) %
\right] }$, where ${\Gamma _{i}=t_{i}^{2}\nu _{Al}}$ and $\nu _{Al}$ is the
density of states in the $Al$. This resonance is only weakly affected by the
superconductivity if ${T_{K}\gg \Delta }$. In this case, the ground state of
the system (impurity plus superconductor) is a singlet and contains even
number of electrons, while the first excited state is localized with the
energy $\epsilon _{K}\lesssim $ ${\Delta }$. In the opposite limit ${%
T_{K}\ll \Delta }$ the trap is always occupied, the ground state is a
doublet and contains odd number of electrons. Because the full Hamiltonian
does not mix states with different parity, as a function of ${T_{K}/\Delta }$
a singlet and doublet level should cross at ${T_{K}^{\ast }\approx \Delta }$%
. Numerical studies confirm this and give the value ${T_{K}^{\ast }\approx
0.3\Delta }$. Close to this value we expect that the energy of the excited
state varies linearly: 
\begin{equation}
\epsilon _{K}={\mathfrak{P}}\,\left( T_{K}-T_{K}^{\ast }\right)
~~~~~~T_{K}^{\ast }\approx 0.3\Delta \;  \label{energybond}
\end{equation}%
with ${{\mathfrak{P}}\sim 1}$. Thus, the electron traps characterized by ${%
T_{K}\sim T_{K}^{\ast }}$ provide a density of localized states at very low
energy. Since the Kondo temperature depends exponentially on the energy $%
\epsilon _{d}^{0}$ of the magnetic impurity, the density of these low energy
states is 
\begin{eqnarray}
\nu (\epsilon ) &=&\rho _{0}V\int_{0}^{\infty }dp(\epsilon _{K})\delta ({%
\epsilon _{K}-\epsilon })\;  \label{denslos} \\
&=&\rho _{0}V\int_{0}^{\infty }\frac{dT_{K}}{T_{K}}\delta \left( {\mathfrak{P%
}}\left( T_{K}-T_{K}^{\ast }\right) \right) =\frac{\rho _{0}V}{{\mathfrak{P}}%
T_{K}^{\ast }}\;  \notag
\end{eqnarray}%
where $\rho _{0}$ is the bare density and $V$ is the volume of the insulator
barrier that contains traps. The density of these states is much larger than
the one of the traps without on-site repulsions. However, their effect on
the qubit is not so large because it is compensated by a small weight of the
Kondo resonance ${w=T_{K}/\epsilon _{d}^{0}}$.

The charge of the qubit produces different electrostatic potentials on the
traps, so electrons/holes tunneling between them causes the qubit
decoherence. We distinguish fast processes in which electrons move a short
distance ${r<\xi }$ which are responsible for high frequency  ($\omega >T$)
behavior of $S(\omega )$, and exponentially slow ones where the distance is
large ${r\gg \xi }$ ($\xi $ is the coherence length of the superconductor).
If the size of the contact, $l$, is small ($l\ll $ $\xi $), the relaxation
might be dominated by tunneling from the traps inside the Josephson barrier
to the one in the surrounding insulator that has much larger volume; however 
for a typical contact $l\sim \xi $ we shall ignore the distinction between
the traps in the barrier and in the insulator nearby. In order to estimate
the contribution of the fast processes, we compare it with the one of TLSs (%
\ref{npsh}). There are two important differences: first, at a given
frequency $\omega $ \textit{any} pair of traps that differ by $\omega $ in
energy can play the role of the two level system if one of these traps is
empty and another is occupied; so the density of states of these pairs is $%
\omega \nu ^{2}(\epsilon )$. Second, the effective dipole moment of the pair
contains the additional small factor, $w$: $\widetilde{p}=wea$ where $a$ is the
typical distance separating the trap from the superconductor. Combining
these factors we get 
\begin{equation}
S_{q}(\omega )=(w\frac{a}{L})^{2}\left( \frac{\rho _{0}V}{{\mathfrak{P}}%
T_{K}^{\ast }}\right) ^{2}e^{2}\omega \;
\end{equation}

The electron tunneling amplitude falls off exponentially for traps separated
by distances larger than $\xi $ leading to the logarithmic distribution of
long relaxation times: ${dP(\tau )=d\tau /\tau }$. Neglecting the
logarithmic dependence of the insulator volume contributing to the slow
dynamics we estimate the number of thermally active pairs by $(T\nu
(\epsilon ))^{2}$. The dipole moment corresponding to this transition is $wea
$, so we get ${S_{q}(\omega )\simeq \frac{\alpha ^{2}}{\omega }}$ with 
\begin{equation}
\alpha \approx \kappa we\left( \frac{\rho _{0}VT}{{\mathfrak{P}}T_{K}^{\ast }%
}\right)   \label{coeff}
\end{equation}%
As expected, the linear frequency dependence at high $\omega $ translates
into a $T^{2}$-dependence of the low frequency power spectrum. Assuming, as
before, that ${V\approx 10^{7}\mathring{A}}^{3}$, ${\rho _{0}\approx 10^{-3}%
\mathring{A}^{3}}$ and ${w\approx 10^{-4}}$ we estimate: ${\alpha \approx
10^{-2}-10^{-3}e}$ for $T\sim 100\;mK$ \cite{lev2}. \ We conclude that these
fluctuators, although very weak, compensate their weakness by their large
number, providing a value for the intensity of the $1/f$ noise in reasonable
agreement with the experiments. Furthermore, they give the correct
dependence of the noise power at low and high frequency. Notice that the
electron moving in and out of the trap has also a large effect on the
Josephson current if this trap is located in the Josephson contact, so these
fluctuators might be responsible also for the noise in the critical current
fluctuations as observed in \cite{Martinis2002}.

In order to estimate the upper cutoff, $\Lambda $, for the $1/f$ noise we
compute the tunneling amplitude between two traps separated by distance $\xi 
$ located in a box of size $\xi $. In the absence of Coulomb repulsion on
each trap, the transition amplitude between two traps with energy $\epsilon $
reads: ${A=\sum_{\alpha }\frac{t_{i}\psi _{\alpha }(i)\psi _{\alpha }(j)t_{j}%
}{\epsilon -\epsilon _{\alpha }}}$. Assuming that the wavefunctions of the
electrons of the superconductor are random we find the tunneling amplitude $%
\overline{|A|^{2}}={\Gamma }^{2}/\nu _{Al}\xi ^{3}\Delta $. The on-site
repulsion changes the energy, $\epsilon $, of each trap to (\ref{energybond}%
) and replaces ${\Gamma \rightarrow }T_{K}^{\ast }$ giving 
\begin{equation}
|A_{0}|^{2}=\frac{T_{K}^{\ast 2}}{\nu _{Al}\xi ^{3}\Delta }=\delta \frac{%
T_{K}^{\ast 2}}{\Delta }\;  \label{transam2}
\end{equation}%
where ${\delta \equiv 1/(\nu _{Al}\xi ^{3})}$ is the average distance
between the levels in the box. For a typical $Al$ island ${A_{0}\approx
10^{-2}T_{K}^{\ast }}$, which is smaller than experimental temperatures. For
a typical aluminum island (${\Delta }_{Al}{\approx 200\mu eV}$) we estimate $%
{A_{0}\approx 10^{8}\;Hz}$. In fact these processes involve energy transfer
to a thermal mode, so  $\Lambda <A_{0}$.

\textit{Conclusion.} We have discussed two possible microscopic mechanisms
responsible for the charge noise in qubits. We found that for realistic
values of the parameters, the conventional TLSs in the insulating Josephson
barrier cannot provide a significant source of decoherence for small qubits,
although they are likely to be the dominant source of noise for large
junctions \cite{Martinis2002}. The TLSs in surrounding insulator might
provide a significant source of noise at ${\omega \lesssim }T\sim 100\;mK$,
but would give frequency independent $S_q(\omega )$ at large $\omega $ and
linear $T-$dependence of the $1/f$ noise. We considered the effect of
Anderson-like traps in the Josephson contact and found that quasiparticle
hopping between these traps provides noise of the right order of magnitude
with the correct temperature and frequency dependence. In order to test this
mechanism one needs to develop a more detailed analytical theory for the
maximal rate of the thermally activated hopping between the traps and for
the effect of the magnetic field that suppresses the superconductivity and
compare these predictions with the data.

\acknowledgments We thank B. L. Altshuler, O. Astafiev, G. Blatter, D.
Geshkenbein, Yu. Paskhin and J. S. Tsai for useful discussions. This work
was supported by NSF DMR-0210575.

\vspace*{-2mm} 
\bibliography{NoiseResub.bbl}

\begin{thebibliography}{24}
\bibitem[{lev()}]{lev2}
\bibinfo{note}{Unlike the situation for the TLSs, the coupling coefficient
  $\kappa=a/L \lesssim1$ for the electrons, because their tunneling amplitude
  is much larger for a given $a$.}
\end{thebibliography}

\end{document}